\documentclass[nofootinbib]{revtex4}
\pdfoutput=1
\usepackage{amssymb,amsmath,graphicx,color}
\begin{document}
\title{Position Space CMB Anomalies from Multi-Stream Inflation}

\author{Yi Wang}
\affiliation{Kavli Institute for the Physics and Mathematics of the Universe (WPI),
Todai Institutes for Advanced Study, University of Tokyo,
5-1-5 Kashiwanoha, Kashiwa, Chiba 277-8583, Japan}

\begin{abstract}
  Temporary domain walls are produced during the bifurcation era of multi-stream inflation. The observational effects from such a domain wall to spectator field perturbations are calculated, and we expect the inflationary perturbations share similarities with the case of spectator field. A domain wall induces a preferred direction in the sky, affecting the angular distribution of perturbations. A correlated suppression of multipoles together with an alignment of multipole moments on the preferred direction are generated. Other observational aspects of multi-stream inflation, including hemispherical asymmetry and cold spot are also briefly reviewed.
\end{abstract}

\maketitle

\section{Introduction}

Recent CMB observations show nice agreements with the simplest inflation model, with nearly Gaussian and nearly scale invariant fluctuations. However, there are still a few anomalies, mainly as position space features, first discussed based on the WMAP data  (\cite{Bennett:2010jb} and references therein) and recently confirmed by Planck \cite{Ade:2013sta} at similar or improved confidence levels. Those anomalies include

\begin{itemize}

\item The angular power spectrum, when locally estimated at different positions on the CMB sphere, are different \cite{Eriksen:2003db, Hansen:2004vq}. Especially the power spectrum on a hemisphere centered at co-latitude $\theta=110^\circ$ and longitude $\phi=237^\circ$ is greater than that on the opposite hemisphere. \footnote{One can alternatively ask if the CMB sky has a dipolar asymmetry in perturbation power spectra \cite{Gordon:2005ai} (see also \cite{Chang:2013vla}).}

\item Looking at more localized scales, a cold spot \cite{Vielva:2003et, Cruz:2004ce} is found on the CMB map. The cold spot
%, centered at $(208^\circ, 304^\circ)$, 
spans about $5^\circ$ in the CMB map and is about $70\mu$K colder than average. Such a spot and the anomalous low temperature is more than $4\sigma$ unlikely from pure statistical chance.

\item In terms of preferred direction for the fluctuation modes, an anomalous alignment of the quadrupole and octopole is also noticed \cite{Tegmark:2003ve, de-OliveiraCosta:2003pu}. The directions for those multipoles coincide with each other up to an error of $9^\circ$, corresponding to an $98\%$ confidence level out of statistical fluctuation (or $13^\circ$ and $96.7\%$ CL from a different analysis).

\item The fluctuations at multiple moment $l\leq 40$ are smaller than expected by $5\%$ to $10\%$, with a statistical significance of $2.5\sigma$ to $3\sigma$ \cite{Planck:2013kta} \footnote{The low $l$ suppression manifests itself in momentum space. Also it is not necessarily related to non-Gaussianity as discussed below. Nevertheless let us list this anomaly together and we will come back to this anomaly in our calculations.}. 

\end{itemize}

Those anomalies characterize some non-Gaussian features of the CMB temperature fluctuations. However, they do not seem to follow directly from nearly scale invariant non-Gaussian correlation functions in momentum space. Because the estimators for non-Gaussianity with nearly scale invariance (see \cite{Wang:2013zva} for a review) are still very consistent with Gaussian. For example, the observational bounds on the local, equilateral and orthogonal shaped non-Gaussianities are \cite{Ade:2013ydc},
\begin{align}
  f_{NL}^\mathrm{local}=2.7 \pm  5.8, \quad
  f_{NL}^\mathrm{equil}= -42 \pm  75, \quad
  f_{NL}^\mathrm{ortho}= -25 \pm  39 \quad \mbox{(68\% CL statistical)}~.
\end{align}
Thus one expects the non-Gaussian features discussed above, if they exist and are not by pure chance, should be either highly scale dependent, or manifest themselves better in position space. It is still unclear if some of those anomalies originate from an incomplete removal of galactic and/or extragalactic foregrounds. Also there is a debate if the anomalies are over-interpreted with too strong posteriori choices.  Nevertheless those anomalies are likely to carry more primordial information than that in the SH initial \cite{Bennett:2010jb}. Here we assume those anomalies comes from primordial physics and search for possible explanations from inflation. 

\begin{figure}[htbp]
  \centering
  \includegraphics[width=0.8\textwidth]{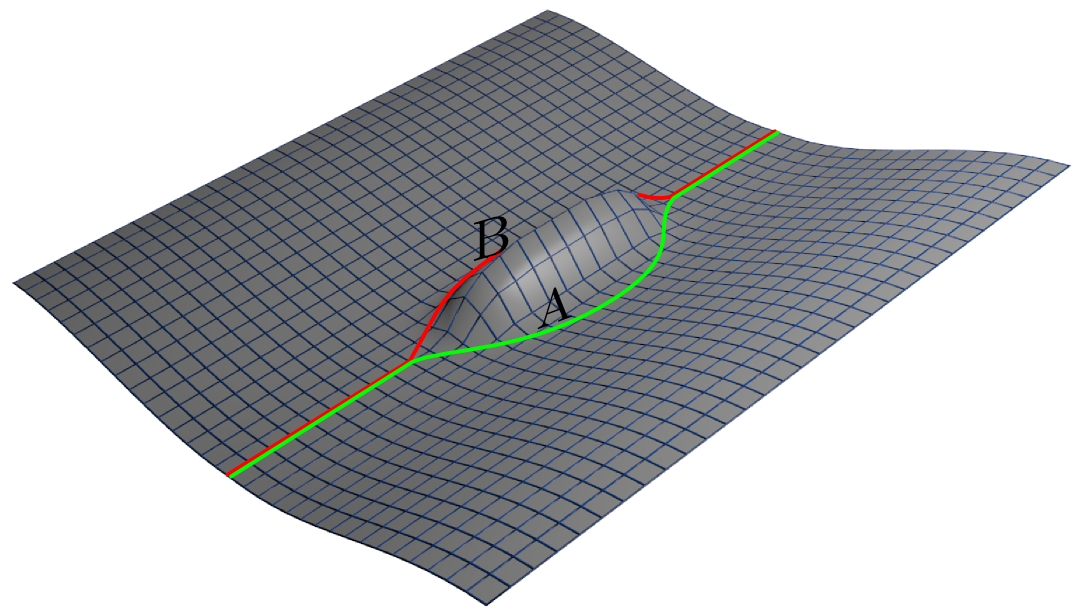}
  \caption{\label{fig:multi-strem} Multi-stream inflation. In this example, the inflation trajectory hit a sharp feature on a two-field potential and has to bifurcate. Different patches of universe follow different trajectories of inflation. The boundaries of those patches are domain walls during the bifurcation era. However, when the trajectories recombine, the tension of the domain walls, and thus the domain walls themselves vanish.}
\end{figure}

The inflation model we focus on is multi-stream inflation \cite{Li:2009sp, Li:2009me, Afshordi:2010wn, Wang:2010rs, Duplessis:2012nb}. During multi-stream inflation, the inflation trajectory hit a bump on the way and bifurcates. Different inflationary local universes, and hence different patches of the CMB sky, follow different trajectories of multi-stream inflation. This situation is illustrated in Fig.~\ref{fig:multi-strem}. Considering that different trajectories are localized at different local universes before they return to the horizon, multi-stream inflation is efficient in producing position space features.

In the remainder of the paper, we shall first review the known features from multi-stream inflation in Sec.~\ref{sec:multi-stre-infl}, and discuss new features from the temporary domain wall in Sec.~\ref{sec:pref-mult-direct} in the context of multi-stream inflation.

\section{Multi-stream inflation}
\label{sec:multi-stre-infl}

The bifurcation behavior in multi-stream inflation can be modeled by different shapes of potentials. For example, the bifurcation can be realized by lowering the potential energy of the preferred trajectories
\begin{align} \label{eq:hybrid-multi-stream}
  V = \frac{m^2}{2} \phi^2  + \frac{\lambda}{4} \left( \chi^2 - \frac{M^2}{\lambda}  \right)^2 + \frac{g^2}{2} \phi^2\chi^2
  + m_1^3 \chi~.
\end{align}
Or alternatively, one can put a Gaussian shaped spike to raise the potential energy of the forbidden region in between the preferred trajectories. This can be achieved by adding
\begin{align}
  \Delta V(\phi,\chi) = M^4 \exp \left( -\xi_1\phi^2 -\xi_2\chi^2 \right) + m_1^3 \chi~.
\end{align}
onto a double field slow roll potential or quasi-single field potential to make the bifurcation. In both examples, a small perturbation is added to allow asymmetry of the two trajectories (the terms $m_1^3 \chi$).

Here we focus on model independent predictions of multi-stream inflation, and shall not discuss the features from departure of slow roll, or field self-interactions during turning of trajectory (see \cite{Abolhasani:2010kn, Abolhasani:2012px} for those discussions based on potential \eqref{eq:hybrid-multi-stream}). The features of multi-stream inflation include \footnote{Some of the features (especially the ones related to the cold spot) correspond to different parameter space of multi-stream inflation. Thus they may not be realized all together from one single bifurcation.}:

\begin{itemize}

\item \textit{Small scale power asymmetry} \cite{Li:2009sp}: When the inflation trajectory bifurcates, the small scale power spectra at different patches of the sky are calculated by perturbation theory around different classical trajectories. Locally, the power spectrum can be still calculated as
\begin{align}
  P_\zeta = \frac{H^2}{8\pi^2M_p^2\epsilon} ~.
\end{align}
However, the local Hubble parameter and slow roll parameter $\epsilon$ can be different along those trajectories. As a result, different patches of the sky admits different small scale power spectra.

For example, when the bifurcation takes place on or before the first e-fold (corresponding to the largest scales) of observable inflation, the asymmetry in the power spectrum is hemispherical. This case is illustrated in Fig.~\ref{fig:avg}.

In multi-stream inflation, the asymmetry in the CMB power spectrum vanishes after (if) different trajectories recombine. The results of Planck show evidence that the hemispherical asymmetry at large multipole moment  $l\gg 100$ may not be present. If it is the case, a recombination of  trajectories at $l\sim 100$ would fit the data.

\item \textit{Average temperature asymmetry} \cite{Li:2009sp}: The total e-folding number along different trajectories could also vary in multi-stream inflation. As a result, different trajectories admit different reheating time and thus different temperature. The difference in temperature follows from a combination of Sachs-Wolfe effect \cite{Sachs:1967er} and the $\delta N$-formalism \cite{Starobinsky:1986fxa, Sasaki:1995aw, Lyth:2005fi}:
\begin{align}
  \Delta T = \frac{1}{5} \Delta N~.
\end{align}

To have this temperature difference under control, it is assumed that the total e-folding number along different trajectories are protected by an approximate symmetry, such that any e-folding number difference between them are highly suppressed. (Otherwise, the additional fluctuation can easily become too large.) This temperature difference is illustrated in Fig.~\ref{fig:avg}.

\begin{figure}[htbp]
  \centering
  \includegraphics[width=0.75\textwidth]{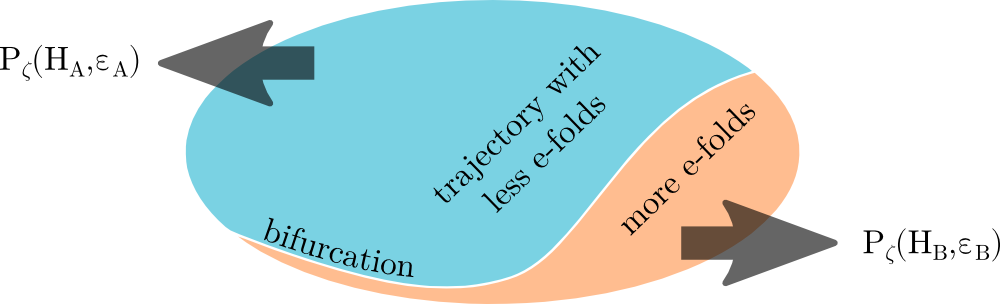}
  \caption{\label{fig:avg} Asymmetries on the CMB induced by multi-stream inflation. The e-folding number difference between the blue and the orange patches brings an averaged temperature difference, whereas the difference in local expansion rate and slow roll parameter brings a difference in small scale power spectra between different patches.}
\end{figure}

\item \textit{Cold spot from last scattering surface} \cite{Afshordi:2010wn}: A cold spot on the CMB may be produced by an extended version of multi-stream inflation\footnote{For other mechanisms to produce the cold spot, see for example \cite{Cruz:2007pe, Bond:2009xx, damo}.}. Note that the cold spot is much smaller than the rest of the CMB sky. Thus the probability for the inflaton to take trajectory A (denoted by $p_A$) or trajetory B (denoted by $p_B$) has to be different. For example, $p_B\ll p_A$.

Let the total e-folds of observational inflation be $N_\mathrm{tot}$, and the bifurcation takes place at the e-fold $N_\mathrm{bif}$ (counted from the end of inflation). Then the radius of the multi-stream bubbles is about $\exp(N_\mathrm{bif}-N_\mathrm{tot})\times 3000$Mpc. The number density of the multi-stream bubble is $p_B \exp(N_\mathrm{tot}-N_\mathrm{bif})$. Again the density contrast is characterized by the e-folding number difference $\Delta N$ between different trajectories. To have a cold spot from the Sachs-Wolfe effect, an over density bubble and thus a smaller number of e-folds is needed inside the bubble. The detailed calculation of the bubble profile can be found in \cite{Afshordi:2010wn}.

\begin{figure}[htbp]
  \centering
  \includegraphics[width=0.35\textwidth]{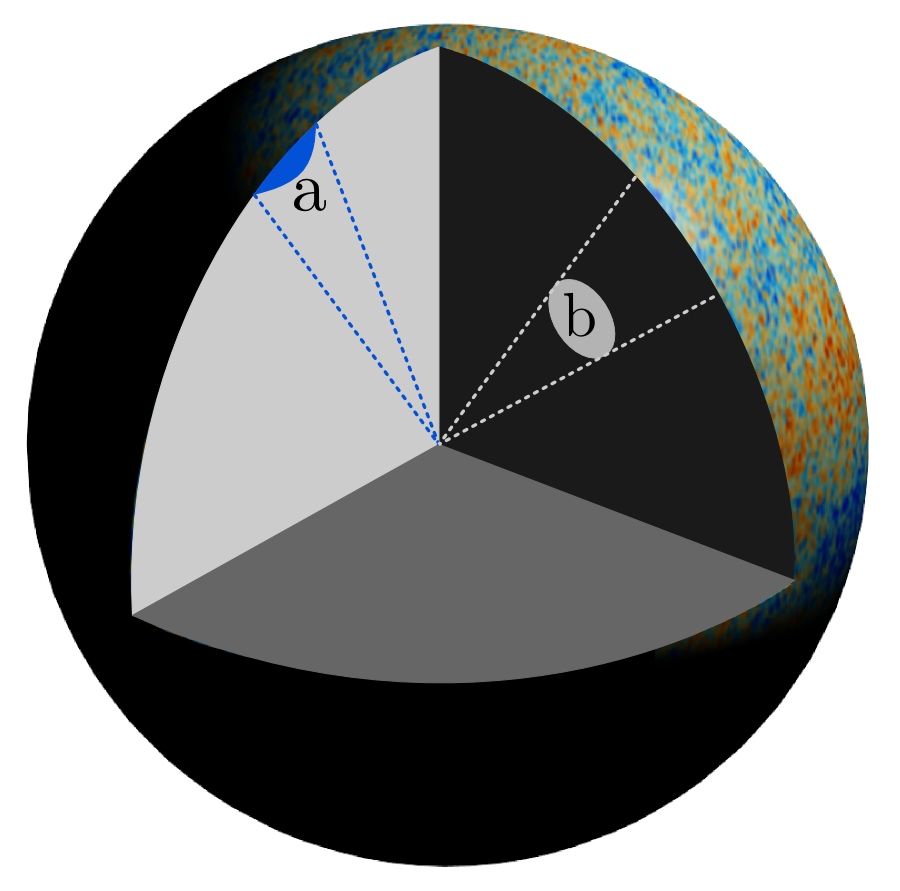}
  \caption{\label{fig:void} Cold spot produced by multi-stream inflation. Either an overdensity region on the last scattering surface (a), or a void in the large scale structure (b), can be produced from a multi-stream bubble. }
\end{figure}

\item \textit{Cold spot from inner void} \cite{Afshordi:2010wn}: In this case, the dominate effects are the  late time ISW effect \cite{Sachs:1967er}, the Rees-Sciama effect \cite{Rees:1968zza} and the Ostriker-Vishniac effect \cite{ostriker}. The qualitative analysis is similar to the above Sachs-Wolfe effect case, but now the multi-stream bubble is located in the large scale structure (for example $z\sim 1$) and a greater number of e-folds is required inside the multi-stream bubble compared with outside.

\item \textit{Constraining multi-field inflation on random potentials}: In case the bifurcations are not under control, multi-stream inflation produce too large anisotropies. Thus those random bifurcations are tightly constrained by current observations. In the case of a two-field random potential, the bifurcation probability is estimated in \cite{Duplessis:2012nb}, which is verified by order of magnitude using numerical simulations. The key observation is that on a random potential, the isocurvature direction has about half chance to be tachyonic. Along the inflationary trajectory, the mass term in the isocurvature direction can be written as a function of the inflaton field value:
\begin{align}
  m_\chi^2 = m_\chi^2(\phi)~.
\end{align}
On a tachyonic potential the isocurvature fluctuations grow exponentially until the mass squared flips sign. There is a phase transition controlled by the averaged frequency of sign flip of $m_\chi^2(\phi)$. On the slower side of the sign flip, the isocurvature fluctuation has long enough time to grow and bifurcation becomes frequent. The bifurcation further enlarges isocurvature fluctuation, and convert the isocurvature fluctuation to curvature fluctuation by the e-folding number difference $\Delta\zeta=\Delta N$ between different trajectories.

On a two field potential, the constraint on the shape of the random potential is not very tight. However, one expects the constraint becomes much tighter (at least with linear growth of bifurcation probability) for a random potential of many fields. The scaling property of bifurcation probability as a function of number of fields is not yet calculated.

\begin{figure}[htbp]
  \centering
  \includegraphics[width=0.55\textwidth]{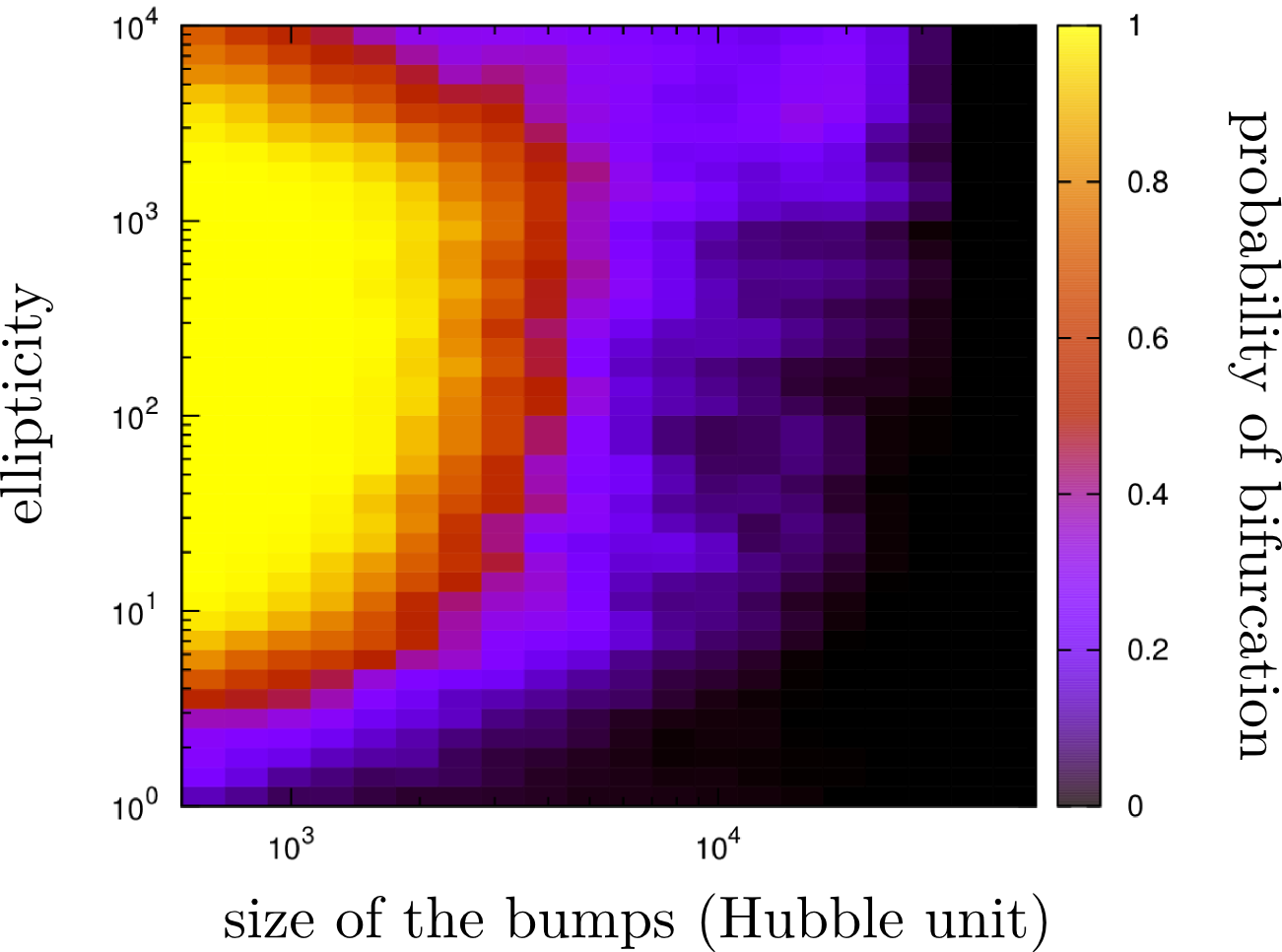}
  \caption{\label{fig:fandom_pot} Bifurcation probability on a two dimensional random potential \cite{Duplessis:2012nb}.}
\end{figure}

\item \textit{Temporary domain wall}: During the bifurcation era (in Fig.~\ref{fig:multi-strem} where the two trajectories are separate), the position space patches following different trajectories are separated by domain walls. If the domain wall is much thinner than the inflationary Hubble radius, the walls themselves do not significantly alter the inflationary background dynamics \cite{Wang:2010rs}. 

For an isocurvature field with potential $V(\chi)=-\frac{1}{2} m_\mathrm{eff}^2 \chi^2+ \frac{1}{4} \lambda \chi^4$, the domain wall has thickness $l \sim 1/m_\mathrm{eff}$ and tension $\sigma \sim m_\mathrm{eff}^3 / \lambda$. Thus we require $m_\mathrm{eff} \gg H$ near the bifurcation \footnote{If this is not satisfied, it remains interesting to see how the domain wall inflates together with the background for a period of time, and what observational features can be produced.}. For the toy potential, $m_\mathrm{eff}^2=M^2-g^2\phi^2$.

The domain wall formation may be fast or slow, depending on the choice of parameters. The $\chi$ field needs to roll a distance $\chi\sim m_\mathrm{eff}/\sqrt{\lambda}$ to form the domain wall. The speed of motion, when $m_\mathrm{eff}\gg H$, can be estimated from $\chi \sim H^2m_\mathrm{eff}^{-1}e^{m_\mathrm{eff} t}$. Thus a quick (i.e. shorter than one Hubble time) formation of domain wall corresponds to $m_\mathrm{eff} < -2 \mathrm{ProductLog}(-\lambda^{1/4}/2)$ or $m_\mathrm{eff} > -2 \mathrm{ProductLog}(-1,-\lambda^{1/4}/2)$. Those solutions are valid for $\lambda < (2/e)^4\simeq 0.29$. The latter solution satisfies $m_\mathrm{eff}>H$ (thin wall). Here we have assumed the inflaton field $\phi$ varies fast such that $m_\mathrm{eff}$ changes quickly enough into a symmetry breaking potential. The parameter regimes for fast and slow bifurcations are plotted in Fig.~\ref{fig:fsb}.

\begin{figure}[htbp]
  \centering
  \includegraphics[width=0.55\textwidth]{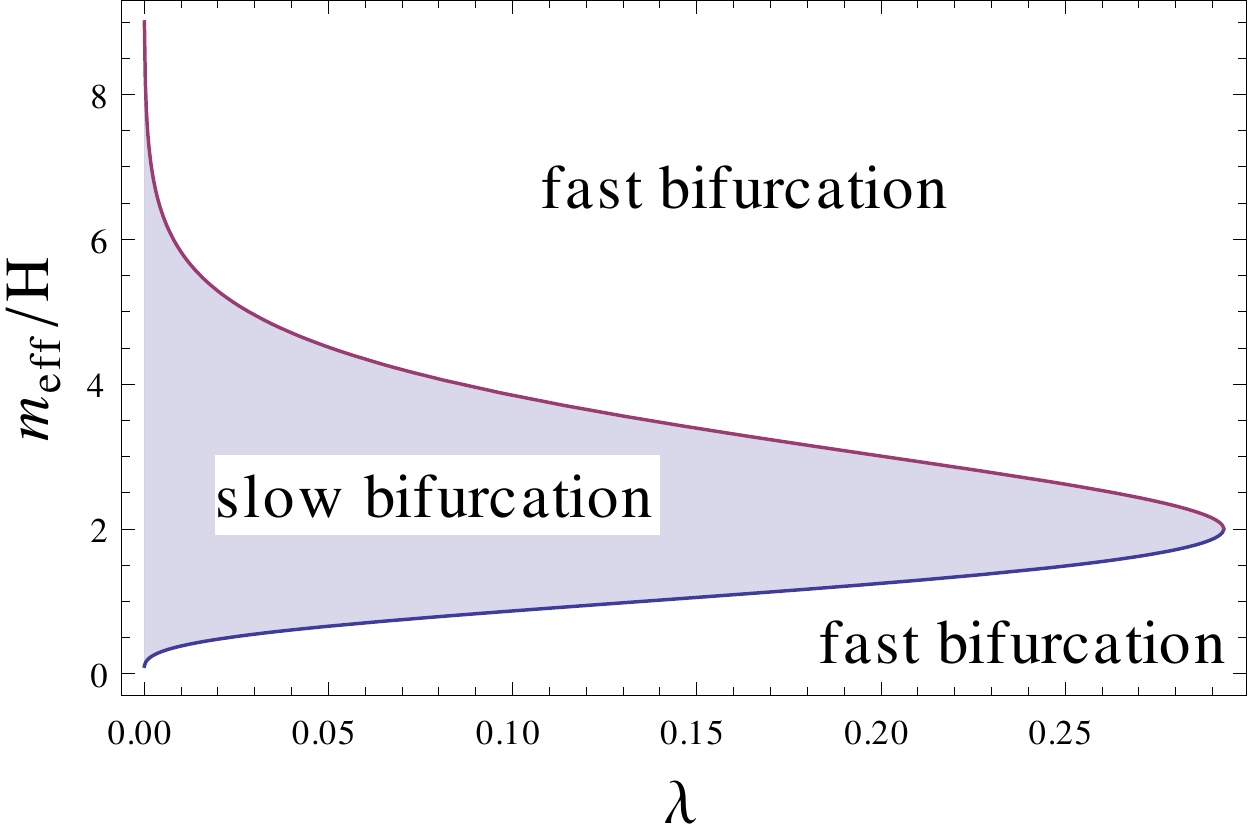}
  \caption{\label{fig:fsb} Fast and slow bifurcations and domain wall formation process. Here we restrict our attention to $\lambda < (2/e)^4\simeq 0.29$ (for $\lambda$ greater than this value, the analytical solution does not apply). The unshaded regime corresponds to fast bifurcation, where the domain wall is formed within one Hubble time. There are two fast bifurcation regimes. The upper one corresponds to a thin wall (compared with Hubble radius). The lower one corresponds to a thicker wall.}
\end{figure}

For the time being, we restrict our attention to a fast bifurcation, with a thin wall, which corresponds to
\begin{align}
  m_\mathrm{eff} > -2 \mathrm{ProductLog}(-1,-\lambda^{1/4}/2)~.
\end{align}
In this case, the domain wall can be considered static during its life time.

Also, those walls disappear when the different inflation trajectories eventually recombine, or reheat into the same thing, or has some rotational symmetry in field space \footnote{We thank  Hassan Firouzjahi for pointing out the third possibility.}. Thus those walls do not cause disasters for late time physics.

However, it is still interesting to see if those temporary domain walls could bring new physics at the level of cosmic perturbations. One step towards this question is taken in Sec.~\ref{sec:pref-mult-direct}.

\end{itemize}

\section{Preferred multipole direction} 
\label{sec:pref-mult-direct}

During multi-stream inflation, a domain wall is temporarily generated between the two classical inflation trajectories. In this section, we consider inflationary perturbations with the presence of a planar domain wall. The validity of this section is not restricted to multi-stream inflation. Most of the calculation also applies for more general inflation scenarios with a thin domain wall. 

\begin{figure}[htbp]
  \centering
  \includegraphics[width=0.4\textwidth]{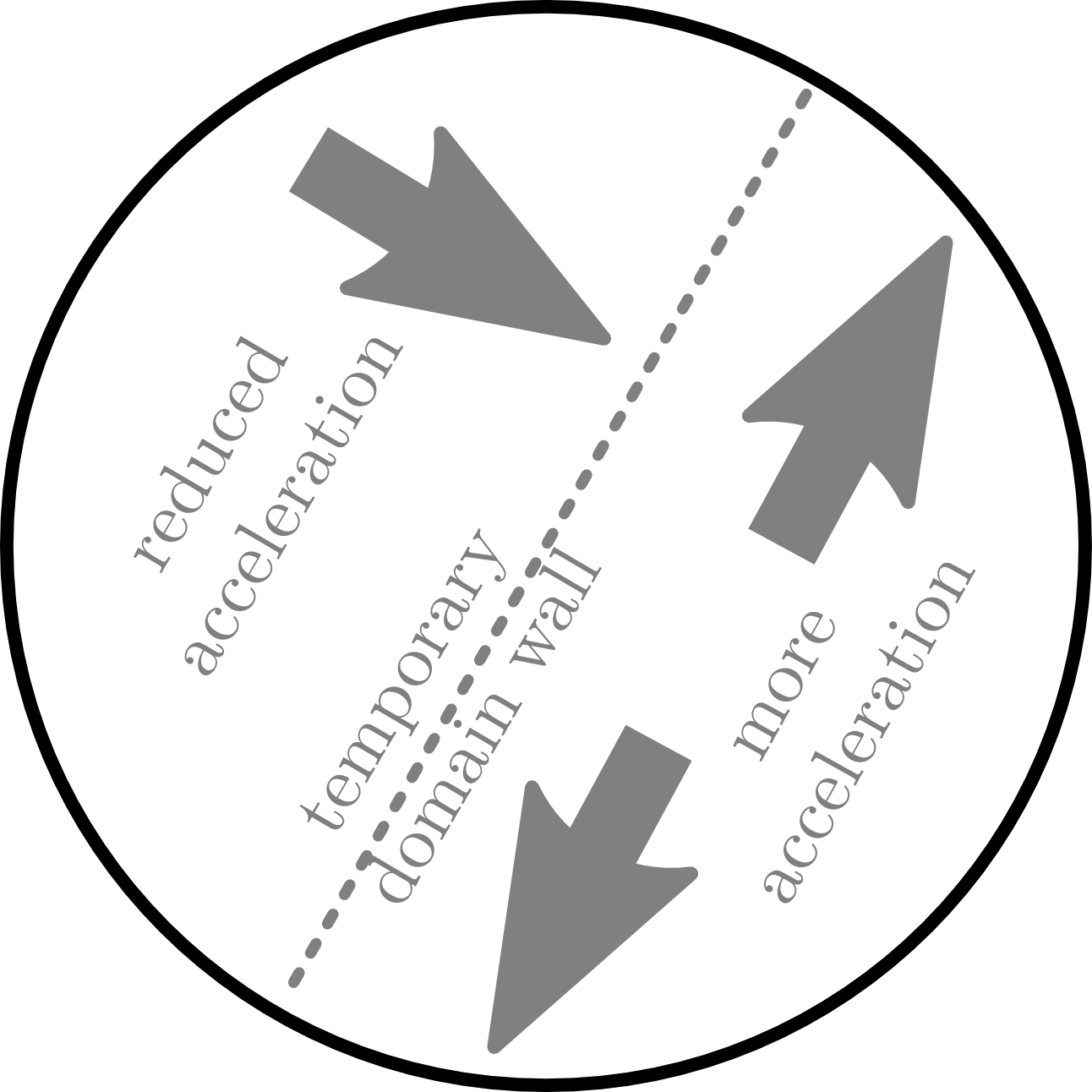}
  \caption{\label{fig:sketch} Additional acceleration during inflation from a domain wall, in the domain-wall-comoving frame. }
\end{figure}

The domain wall solution can be understood in different frames. Here we consider two frames namely the domain-wall-comoving frame and the observer-comoving frame. In the domain-wall-comoving frame, with the domain wall located at $z=0$, the stress tensor of the domain wall can be written as
\begin{align}
  T^\mu_\nu = \Lambda M_p^2  ~\mathrm{diag}(1,-1,-1,-1) + \sigma \delta(z) ~\mathrm{diag}(1,-1,-1,0)  ~,
\end{align}
where $\Lambda$ is the effective cosmological constant that drives inflation, and $\sigma$ is the tension of the domain wall. The Einstein equations with the above stress tensor can be solved by \cite{Wang:2010mb, Wang:2011pb} (see also \cite{Bowcock:2000cq}) \footnote{Here we consider the planar solution. Domain wall solutions with alternative topology is known in \cite{Cvetic:1993xe}, which follows more closely from \cite{Vilenkin:1981zs}. }
\begin{align} \label{eq:dw1}
  ds^2 = \frac{1}{\alpha^2(\eta + \beta|z|)^2} \left( -d\eta^2 + dz^2 + dx^2 + dy^2 \right)~,
\end{align}
where
\begin{align}
  \alpha \equiv \sqrt{\frac{\Lambda\Gamma}{12}} (\Gamma+1)~, \quad \beta \equiv \frac{\Gamma-1}{\Gamma+1} ~, 
\end{align}
\begin{align}
  \Gamma \equiv 1+ \frac{3\varepsilon - \sqrt{48\varepsilon+9\varepsilon^2}}{8} ~, 
  \quad \varepsilon \equiv \frac{\sigma^2}{M_p^4 \Lambda} ~.
\end{align}
Note that $\beta$ takes value in the range $-1<\beta\leq 0$. This solution is illustrated in Fig.~\ref{fig:sketch}, that the acceleration in the $z$ direction is reduced compared with a purely inflationary background. As we shall see, the reduced acceleration corresponds to a suppression of fluctuation along the $z$ direction.

This solution can be understood in Fig.~\ref{fig:sketch}, that in the $z$ direction, which is perpendicular to the direction of the wall, the acceleration is reduced compared with that on the $x$-$y$ plane. Note that this statement is observer dependent. In the observer-comoving frame, we get a different conclusion. Nevertheless, this domain-wall-comoving frame relates more closely to the observations.

Alternatively, one can do a coordinate transformation to get to the observer-comoving frame. When $z>0$, the metric can be written as  \cite{Wang:2010mb, Wang:2011pb}.
\begin{align}\label{eq:dw2}
  ds^2 = \frac{3}{\Lambda \check \eta^2} \left( -d\check\eta^2 + d\check z^2 + d\check x^2 + d\check y^2\right)~,
\end{align}
with the coordinate transformation
\begin{align}
  \check\eta = \gamma (\eta+\beta z) ~, \quad \check z = \gamma(z+\beta\eta)~,
  \quad \check x = x~, \quad \check y = y~,
\end{align}
where $\gamma\equiv 1/\sqrt{1-\beta^2}$. Thus $\beta$ behaves as a Lorentz boost between the observer-comoving frame \eqref{eq:dw2}, and the domain-wall-comoving frame where the domain wall is static \eqref{eq:dw1}. In the coordinate \eqref{eq:dw2}, the position of the domain wall is $\check z = \beta\gamma\eta$. Thus  the domain wall is moving away from the observer. The metric in the $z<0$ part can be written down similarly, with a boost in the opposite direction $\beta\rightarrow-\beta$.

Now let us consider how such a domain wall affects the fluctuations, which originate from sub-Hubble scales. We consider a spectator scalar field $\delta\phi$ gravitationally coupled to the domain wall. Using coordinate \eqref{eq:dw2}, the calculation is standard, with
\begin{align}
  \delta \check \phi (\check x) = \int \frac{d^3 k}{(2\pi)^3} e^{i \mathbf{k}\cdot\check{\mathbf{x}}} 
  u_k(\check \eta) a_\mathbf{k} + \mathrm{h.c.}~,
\end{align}
where
\begin{align}
  [a_\mathbf{k},a^\dagger_{\mathbf{k}'}] = (2\pi)^3 \delta^3(\mathbf{k}-\mathbf{k}')~,
\end{align}
and
\begin{align}
  u_k(\check \eta) = \frac{H}{\sqrt{2k^3}} (1+i k \check \eta) e^{-i k \check \eta}~.
\end{align}
Here the initial condition of the perturbation is chosen such that the perturbation \textit{locally} minimize the energy density on the $z>0$ patch. This coincides with the Bunch-Davies vacuum on this patch. Also, note that the horizon crossing time is controlled by $\check \eta$ instead of $\eta$.

To relate to observations \footnote{Here we assume the domain wall (before it disappears) is located crossing the center of our observable universe. Thus the domain wall rest frame is preferred by the symmetry of the system.}, we have to transform to frame \eqref{eq:dw1}. On super-Hubble scales ($-k \check \eta\ll 1$), the transformation can be written as
\begin{align}
  \delta\phi(x) & = \delta\check \phi (\check x) = \int \frac{d^3 k}{(2\pi)^3} e^{i[k_1 x + k_2 y + k_3(z+\beta\eta)\gamma]}
  \frac{H}{\sqrt{2k^3}} a_\mathbf{k} 
  + \mathrm{h.c.} \nonumber\\ &
  = \int \frac{d^3 q}{(2\pi)^3} e^{i \mathbf{q}\cdot{\mathbf{x}}} 
  \left[ q_1^2 + q_2^2 + \gamma^2q_3^2  \right]^{-\frac{3}{4} } \frac{\gamma H}{\sqrt2}  a_{\mathbf{k}(\mathbf{q})} + \mathrm{h.c.} ~,
\end{align}
where we have used that $\check \eta \simeq 0$, and thus $\eta\simeq -\beta z$. Note that the equal $\check \eta$ hypersurfaces should be used here because it is $\check \eta$ that controls the horizon crossing and frozen behavior . In the second line, we have changed the integration variable into $q$ such that the different spatial directions have different weight: $k_1=q_1$, $k_2=q_2$ and $k_3 =\gamma q_3$. Thus the Fourier space correlation function can be calculated as
\begin{align} \label{eq:2pt}
  \langle\delta\phi_\mathbf{q}\delta\phi_{\mathbf{q}'}\rangle = (2\pi)^3 \delta^3(\mathbf{q}-\mathbf{q}') \frac{\gamma H^2}{2}  \left( q_1^2+q_2^2+\gamma^2q_3^2 \right)^{-\frac{3}{2}}~,
\end{align}
where we have used $\delta^3(\mathbf{k}-\mathbf{k}') = \gamma^{-1} \delta^3(\mathbf{q}-\mathbf{q}')$. Note that the result is symmetric under $\beta\rightarrow-\beta$. Thus this correlation function can also be extended into the $z<0$ regime.  Clearly, the two point correlation function \eqref{eq:2pt} has a preferred direction, perpendicular to the domain wall. 

% In the language of background field equations, the domain wall is located at $\chi=0$, along the inflaton $\phi$'s motion direction and not moving by itself. Thus in order to use the inflaton as the clock field, the domain wall rest frame should be used. \footnote{The same conclusion can be made from arguments of symmetry. The physics on different sides of the domain wall are symmetric. Especially, in this paper we focus on the case where the domain wall divide the CMB sky into two symmetric hemispheres. When we make observations on both sides of the wall, there is no preferred direction of motion for the domain wall.} 

% After identified the clock field, the $\delta N$ formalism can be applied to verify the time delay formula $\zeta = - H \delta\phi/\dot\phi$. As a result, the power spectrum for $\zeta$ is now \footnote{Note that the time dependence of the inflaton background $\phi$, together with the gravitational perturbations are not included in the calculation here. We expect the local calculation around horizon crossing not being sensitive to those complexities. On the other hand,  a curvaton scenario \cite{Linde:1996gt, Enqvist:2001zp, Lyth:2001nq} can be considered to by pass the gravitational perturbations. In the case of a curvaton, the angular dependence of perturbations is not modified. Only an overall factor in front of $P_\zeta$ is changed because of different conversion mechanism from isocurvature to curvature.}

Now we would like to discuss the conversion from this spectator perturbation to inflationary curvature  perturbation $\zeta$. The conversion could occur via the curvaton mechanism \cite{Linde:1996gt, Enqvist:2001zp, Lyth:2001nq} or the modulated reheating \cite{Kofman:2003nx, Dvali:2003ar, Suyama:2007bg}. In those cases, the spectator field does not have a moving background thus the conversion from the spectator perturbation to the inflationary curvature perturbation is linear. 

On the other hand, it would be more interesting to interpret the spectator field $\delta\phi$ as the perturbation of a rolling inflaton field $\phi = \phi_0 + \delta\phi$. However, to do this, one has to specify how to match the background value of $\phi_0$ on the spacetime hypersurfaces before/after the domain wall has formed/disappeared, considering that $\phi_0$ is time dependent. We hope to address those matching conditions in future work. On the other hand, it is intuitive to assume that the non-trivial matching and dynamics of $\phi_0$ eventually get diluted after the domain wall disappeared. Only the anisotropic quantum fluctuations $\delta\phi$ survives because those fluctuations are frozen on super-Hubble scales. 

Thus we have linear conversion, with constant coefficient, from $\delta\phi$ to $\zeta$ in curvaton or modulated reheating scenarios, and we assume linear conversion when interpreting $\delta\phi$ as the inflaton fluctuations. Inspired by the $\delta N$ formalism for the inflaton fluctuations, here we use the conversion factor $\zeta = - H \delta\phi/\dot\phi$ in the reminder of this section. For curvaton or modulated reheating, this overall conversion factor is modified but no other discussions are to be changed.

The power spectrum is now
\begin{align}
  P_\zeta (\mathrm{q}) =  \gamma \left(\frac{q^2}{q_1^2+q_2^2+\gamma^2q_3^2}\right)^{\frac{3}{2} } \frac{H^2}{8\pi^2M_p^2 \epsilon} ~.
\end{align}

\begin{figure}[htbp]
  \centering
  \includegraphics[width=0.8\textwidth]{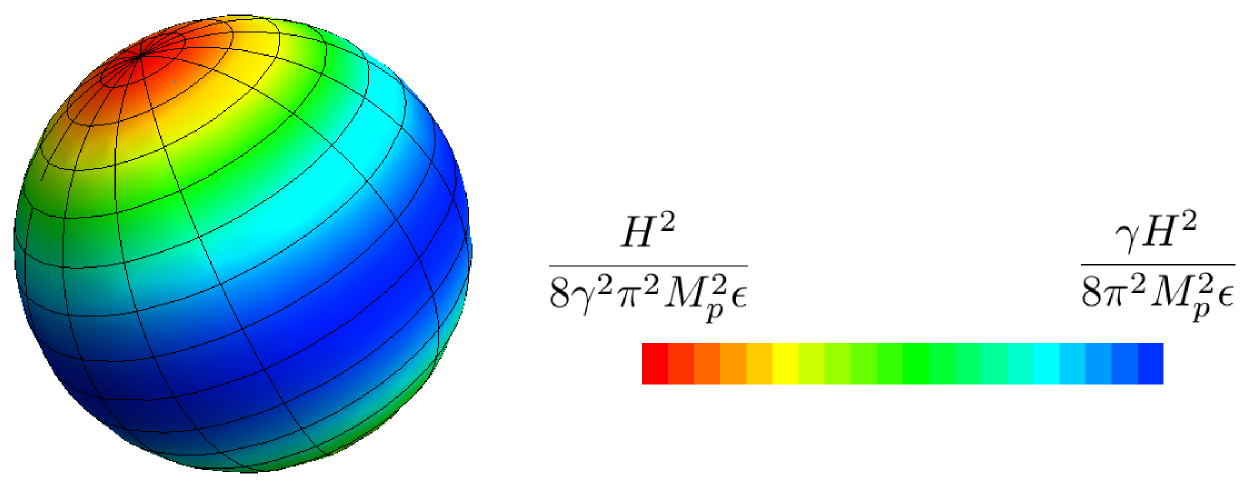}
  \caption{\label{fig:power-dist} The power spectrum with a domain wall, along different directions. The domain wall is located at the equator. Note that this plot shows the preferred direction for the fluctuation power spectrum, not the temperature fluctuation itself. The suppressed power spectrum (in red color) is more probable to show up. This can be related to the lack of power for the first a few multipoles.}
\end{figure}

Note that there the Lorentz boost factor $\gamma > 1$. Thus the probability to find the fluctuation along the $q_3$ direction is suppressed compared with that on the other two directions. This angular dependence is plotted in Fig.~\ref{fig:power-dist}.

% \begin{align}
%   \frac{H^2}{8\gamma^2\pi^2M_p^2 \epsilon} \qquad \frac{\gamma H^2}{8\pi^2M_p^2 \epsilon}
% \end{align}

When averaged over the angular directions, the power spectrum is
\begin{align}
  P_\zeta (q) = \int  \frac{d\theta ~\sin\theta}{2} P_\zeta (\mathrm{q}) = \frac{H^2}{8\pi^2M_p^2 \epsilon}~.
\end{align}
Thus the angular-averaged power spectrum is not modified. Only that a preferred direction is generated. However, this angular average of the power spectrum is naive, due to the existence of a preferred direction.
 
The power spectrum itself, even as a function of vector $\mathrm{q}$, does not fully characterize the statistical properties of perturbations. Because the power spectrum only shows the size of fluctuation given one direction, but does not show a difference in probability to get fluctuations pointing to different directions.

\begin{figure}[htbp]
  \centering
  \includegraphics[width=0.4\textwidth]{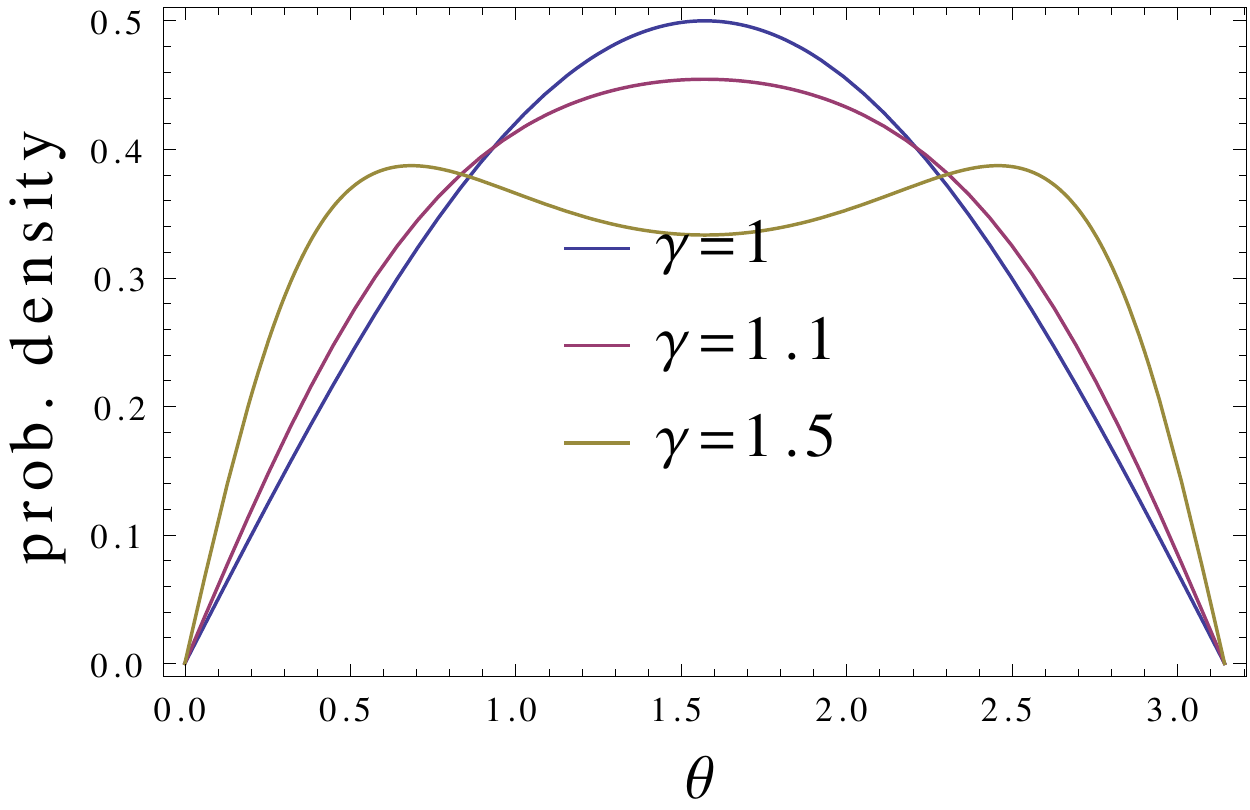}
  \hspace{0.1\textwidth}
  \includegraphics[width=0.4\textwidth]{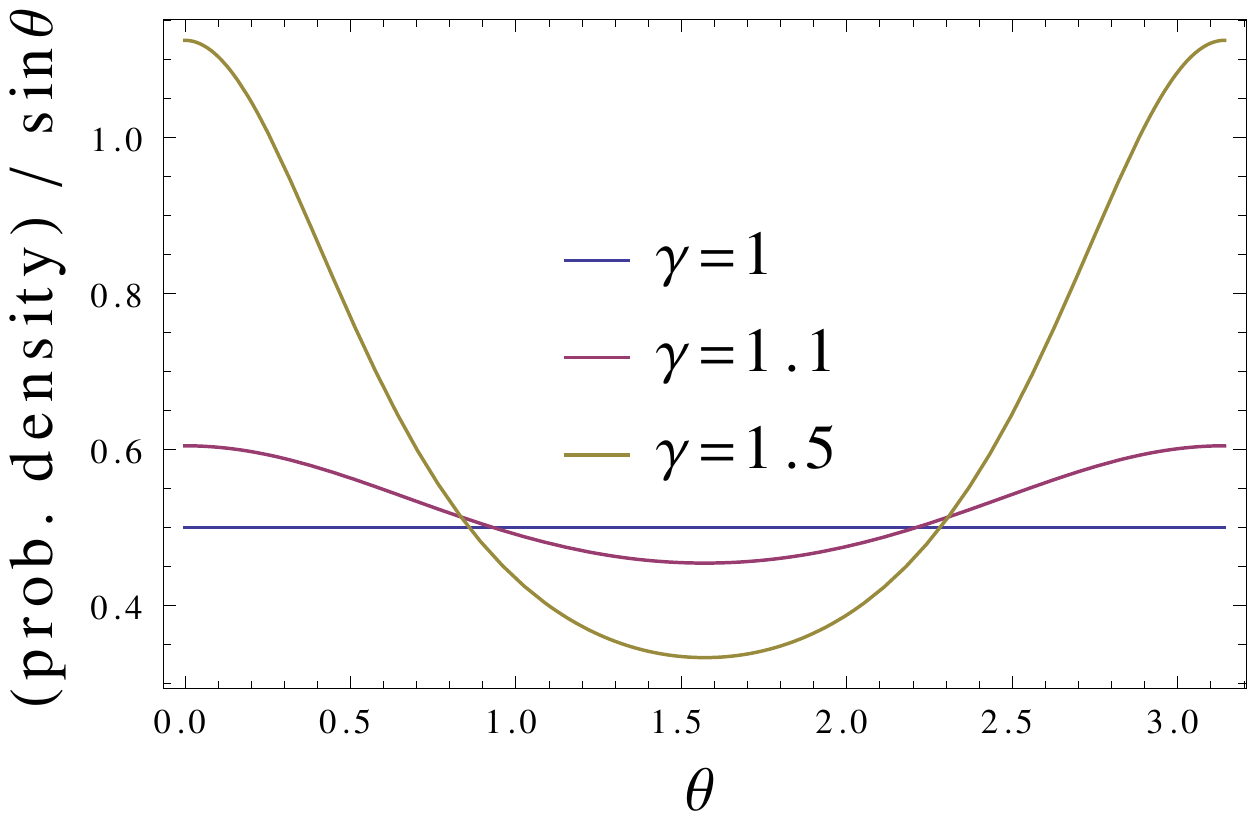}
  \caption{\label{fig:dist} Probability distribution function of direction angle $\theta$ for different $\gamma$ factors. The $\gamma=1$ choice corresponds to the case without a domain wall.}
\end{figure}

In the case with a domain wall, this preferred direction does exist. This is because the effect from a domain wall corresponds to a Lorentz boost, and a Lorentz boost transforms a sphere into an ellipse. In the observer-comoving frame, there is no preferred direction which has a greater probability to observe fluctuations. In the domain-wall-comoving frame, the $\beta$ factor brings a preferred direction. To find the preferred direction and corresponding probability distribution, let us consider points on the $x$-$z$ plane. Using polar coordinate in both frames, we have
\begin{align}
  x = R \sin \theta = \check R \sin \check \theta~, \qquad z = R \cos \theta = \gamma \check R \cos \check \theta~.
\end{align}
Again the mapping is on equal $\check \eta$ hypersurface, because such hypersurfaces control horizon crossing scales. Here $\check \theta$ has probability density satisfying distribution 
\begin{align}
  \mathrm{Prob}(\check \theta) = \frac{1}{2}  \sin \check \theta~.
\end{align}
The probability distribution of $\theta$ is determined by
\begin{align} \label{eq:theta-prob}
  \theta = \arctan \left( \frac{1}{\gamma} \tan\check\theta \right)~, \qquad \mathrm{Prob}(\theta) = \frac{\gamma^2 |\tan\theta|}{2\cos^2\theta(1+\gamma^2\tan^2\theta)^{\frac{3}{2} }} ~.
\end{align}
The probability distributions are plotted in Fig.~\ref{fig:dist}. Note that the $\theta$ angle here is the direction along which the perturbation propagates. The orientations of the multiples (defined in \cite{de-OliveiraCosta:2003pu, Ade:2013sta}) point to the perpendicular direction to $\theta$, which peak at directions on the domain wall. On the other hand, the $\phi$ direction still has a uniform distribution, following from the symmetry of the configuration.

It is interesting to note that the directions with $\theta$ close to $0$ and $\pi$ are more probable. Those preferred directions have suppressed power spectra, as plotted in Fig.~\ref{fig:power-dist}. Thus it is interesting to see if the preferred direction is related to a small-$l$ suppression of the power spectrum \footnote{In multi-stream inflation, in case that the inflation trajectories recombines at a later time, the higher $l$ multipoles are no longer suppressed.}. 

\begin{figure}[htbp]
  \centering
  \includegraphics[width=0.55\textwidth]{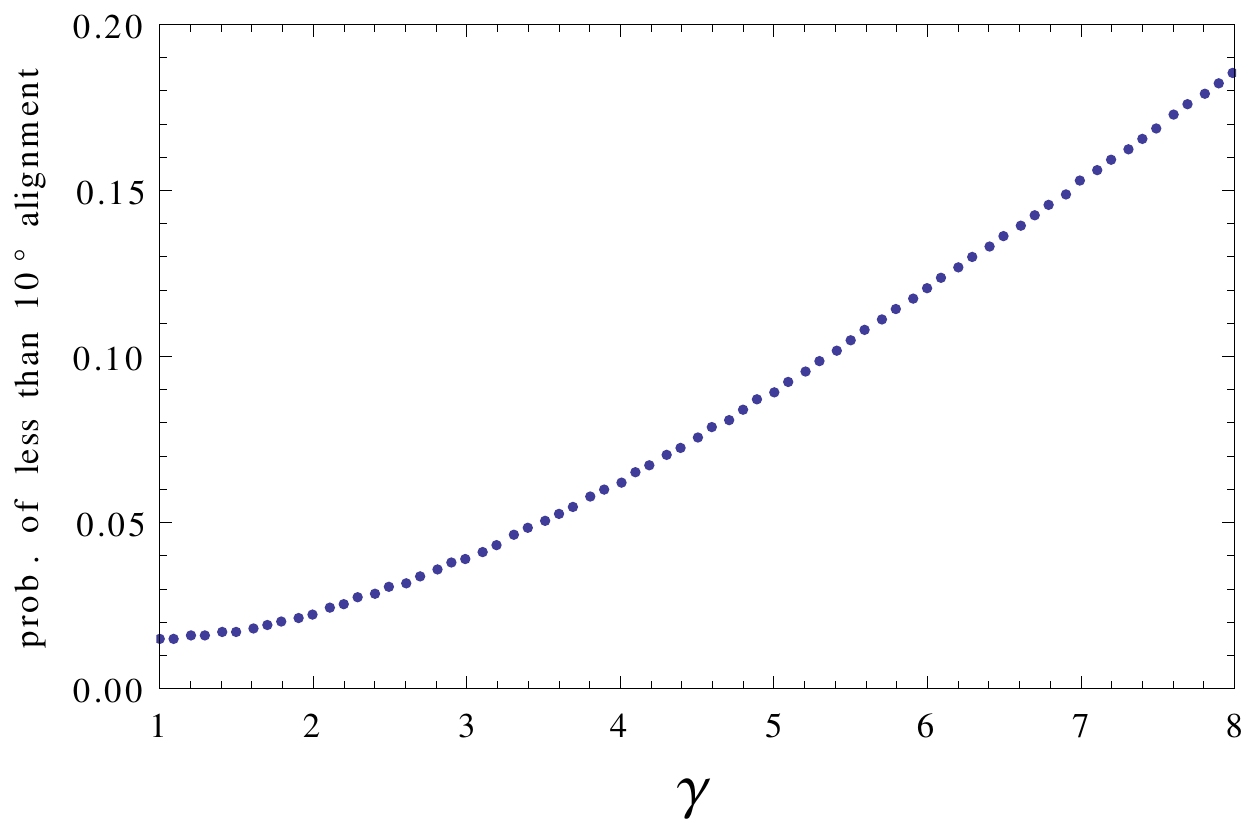}
  \caption{\label{fig:align} Probability of alignment for two multipoles. The probability is calculated numerically based on $10^8$ random samples per $\gamma$ value. Here for each sample, we generate a pair of random directions in 3-dimensional space obeying the probability distribution \eqref{eq:theta-prob}, and calculate the probability that those two random directions has a relative angle smaller than 10$^\circ$. This is similar to the angle defined in \cite{Ade:2013sta}. However, in \cite{Ade:2013sta} the angle is defined from directions of $a_{lm}(\mathbf{\hat n})$ while we define the angle with respect to the 3-dimensional momentum $\mathbf{k}$. We hope to consider the 3-dimension to 2-dimension projection, together with more realistic comparison with the alignment angle in the future. Note that the probability of alignment becomes much higher for large $\gamma$. }
\end{figure}

Based on the probability distribution function, one can further check the probability of mode alignments. The angle between two vectors $(\theta_1,\phi_1)$ and $(\theta_2,\phi_2)$ is
\begin{align}
  \delta = \arccos \left[  \cos\theta_1\cos\theta_2 + \sin\theta_1\sin\theta_2\cos(\phi_2-\phi_1)\right]~.
\end{align}
The probability for two modes with direction vectors lying within $\delta < 10^\circ$, as a function of $\gamma$ is plotted in Fig.~\ref{fig:align}. Note that here we do not map the direction vector to the two dimensional CMB sphere. Thus the probabilities are lower than the numbers listed in \cite{Ade:2013sta}. The figure shows an increase of probability with increase of  $\gamma$. Note that if one takes large $\gamma$, the multipoles are suppressed correspondingly. It remains interesting to see if such a correlated suppression of low multipole moments and alignment better fit the CMB data.

Before ending up this section, a few comments are in order here.
\begin{itemize}
\item We have matched observational quantities with the quantities in the rest frame of the domain wall. The calculation is done via a boost from the frame \eqref{eq:dw2}. Thus there is a question if such a boost produce different Doppler dipoles on different sides of the domain wall. Such dipoles, if exist, induce huge perturbations and thus constrain $\beta$ strongly. However, one should note that the physical clock field during inflation is not boosted by the domain wall (at least in the case of multi-stream inflation). Thus there is no anisotropy from the boost of the background. On the other hand, the perturbations crosses the horizon following a boosted clock. Nearly scale invariance of the perturbations protects the consistence of the above picture.
\item In \cite{Wang:2011pb}, the cosmological perturbations are also calculated. However, \cite{Wang:2011pb} focus on a different situation, in which the whole observable universe is on one side of the domain wall. Thus the observables considered in \cite{Wang:2011pb} are different from the ones considered here. As a related topic, the cosmological perturbation of the domain wall itself is considered in \cite{Garriga:1991ts}.
\end{itemize}

\section{Conclusion and discussions}

To conclude, we consider how domain wall affects the amplitude and angular distribution of inflationary fluctuations. The domain wall generates a Lorentz boost. The boost does not modify the angular-averaged power spectrum of perturbations from a naive average. However, the fluctuations along the direction perpendicular to the domain wall is suppressed by $1/\gamma^2$, and the fluctuations along the directions parallel to the domain wall is amplified by $\gamma$.

Due to the boost, the density fluctuation becomes more probable to propagate along the direction perpendicular to the wall (Fig.~\ref{fig:dist}). The preferred direction reduces the fine tuning of mode alignment (Fig.~\ref{fig:align}, which affects for example the alignment of the quadrupole direction with the octopole direction). This alignment is correlated with a suppression in power spectrum. It is interesting to further investigate how much better those correlated effects fit the CMB data.

In the original model of multi-stream inflation, the domain wall exists only temporarily. The different inflation trajectories eventually recombine and the domain wall decays during inflation. We expect that after the decay of the domain wall, the domain wall energy density gets diluted quickly. Thus we do not observe a wall-like over density regime on cosmological scales (the opposite possibility is discussed in \cite{Li:2009me}). 

It would be interesting to consider the true time dependence of the domain wall, which may correspond to a time dependent boost factor. Intuitively after the perturbations become classical and frozen on super-Hubble scales, the vanishing tension of the domain wall no longer modifies the frozen perturbations. On the other hand, the smaller scale modes which exits the horizon after the disappearance of the domain wall should remain standard and unmodified. Only perturbations which exit the horizon during the bifurcation era are affected. We hope to address this issue in future work. 

In the calculation a spectator field is considered thus strictly speaking one still need to consider the conversion from this field fluctuation to curvature perturbation. We expect the calculation can be directly applied to the inflaton, considering the perturbation of the inflaton should behave like a spectator field on sub-Hubble and Hubble crossing scales in the $\delta\phi$-gauge (as in the case of the standard cosmic perturbation theory). This assumption needs to be verified.

Also, only the planar class of domain wall solutions is considered here. It remains to verify if multi-stream inflation approximately generates the planar domain wall (which looks intuitive), or other classes of domain wall solutions are more preferred.

Finally, the essential mechanism discussed here is the boost generated by the domain wall. It is interesting to consider if there are mechanisms other than from a domain wall, which generate a similar boost and thus similar predictions.

\section*{Acknowledgments}
We thank Xingang Chen, A.Emir Gumrukcuoglu, Chunshan Lin, Yin-Zhe Ma and Shinji Mukohyama for discussion. The research is supported by fundings from the Kavli Institute for the Physics and Mathematics of the Universe. 

% Generated by or.py

\end{document}